\documentclass[11pt]{article}

\usepackage{epsfig}
\usepackage{cite}

\begin{document}
\setcounter{page}{1}

\begin{center}
\begin{Large}
{Spin physics with STAR}  \\
\end{Large}

\vspace*{0.2in} 
Joanna Kiryluk for the STAR Collaboration\\
\vspace*{0.1cm}
{\footnotesize{ {\em Department of Physics and Astronomy, University of California Los Angeles, \\ 
 Los Angeles, California 90095-1547 USA \\ }
}}
\end{center}

\vspace*{0.2in}

\begin{abstract}
The  STAR collaboration aims to study polarized proton-proton collisions
at RHIC.
The  emphasis  of  the  spin  run this year is on transverse single spin
asymmetries.
Beyond  2001,  we  aim   to  determine directly  and precisely the gluon
polarization, as well as the polarizations of the  $u$,  $\bar{u}$,  $d$
and $\bar{d}$ quarks in the proton by measuring in addition longitudinal
and double spin asymmetries.
Furthermore, we aim to measure for the first time the quark transversity
distributions.
These  measurements  will  improve  substantially  the   knowledge   and
understanding of the spin structure of the nucleon.
\end{abstract}

\section{Introduction}    %) A SECTION HEADING
Over  the  past  decade,  the inclusive spin asymmetries $A_{1(2)}$ and
structure functions $g_{1(2)}$ of the nucleon have been measured in Deep
Inelastic Scattering (DIS)  experiments  at  CERN\cite{emc,smc},  
SLAC\cite{slac} and DESY\cite{desy}.
The  data lead one to conclude that only a small fraction of the nucleon
spin is carried by quark spins.
The  gluon  polarization  is  determined indirectly from next-to-leading
order QCD analyses of the scale dependence of the data, and is found  to
be relatively small to within large uncertainties\cite{smc}.
Semi-inclusive  measurements\cite{smc_semi,hermes_semi} demonstrate that the
polarization of the valence $u$  quark  is  positive  and  that  
the polarization of the valence $d$ quark is negative.
The  data  do not resolve the flavor composition of quark spin densities
in the sea.
Only under the assumption of  SU(2) flavor/isospin symmetry of the light
sea  is  the polarization of the sea quarks/anti-quarks determined,  and
found consistent with zero to within large experimental uncertainties.

The largely open questions of which fraction of the nucleon spin originates
from  gluon  polarization  and which fraction results from sea-quark
polarization will  be  addressed  in  the  present  generation  of  spin
experiments, including STAR at the Relativistic Heavy Ion Collider (RHIC).

For this purpose, RHIC will accelerate and collide intense beams
($\mathcal{L} = 8 \times 10^{31}$ -- $2 \times 10^{32}\,\mathrm{cm}^{-2}\mathrm{s}^{-1}$)
of highly polarized (70\%) protons, to projected integrated luminosities
of  \mbox{320} \\ 
\mbox{(800)\,pb$^{-1}$}  at   \mbox{$\sqrt{s}=$200\,(500)\,GeV}
over the upcoming years\cite{bunce}.

STAR (Solenoid Tracker At RHIC) will  measure  the  cross-section  asymmetries  in collisions with
either beam or both beams  polarized in longitudinal ($L$) and tranverse
($T$) directions.
These so-called single and double spin asymmetries can be denoted by:
\begin{equation}
  A_{L,T} =  {{1}\over{P}} \cdot
             {{ R_{-} - R_{+} } \over
                    { R_{-} + R_{+} }}, \hspace{0.1in}
  A_{LL,LT,TT} = {{1}\over{P_1 \, P_2}} \cdot
                 {{ (R_{++} + R_{--}) - (R_{+-} + R_{-+}) } \over
                  { (R_{++} + R_{--}) + (R_{+-} + R_{-+}) }},
  \label{asymmetries}
\end{equation}
where  the  symbols  $P$  denote  the respective beam polarizations, the
symbols $R$ denote the observed  number  of  events  normalized  to  the
luminosity   of  the  beam  crossing, and the subscripts $(+)$ and $(-)$
denote the spin directions with respect to the $L,T$ axes.
These measurements thus require knowledge of the relative luminosities for
$(+)$ versus $(-)$, and for $(++)$ and $(--)$  versus  $(+-)$ and $(-+)$
collisions,   in   addition   to   knowledge   of   the   absolute  beam
polarizations.

The asymmetries give insight into the nucleon spin structure.
Specifically, STAR will study:
\begin{enumerate}
\item{the gluon polarization from the measurements of $A_{LL}$
      (i)   in inclusive high $p_T$ (prompt) photon production
            $\vec{p}\vec{p} \rightarrow \gamma + X$,
      (ii)  in jet(s) production $\vec{p}\vec{p} \rightarrow {\rm jet(s)} + X$,
      (iii) from the prompt photon in coincidence with jet production
            $\vec{p}\vec{p} \rightarrow \gamma + {\rm  jet} + X $,
            and
      (iv)  possibly from heavy quark production
            $\vec{p}\vec{p} \rightarrow e^{+} + e^{-} +X$,
}
\item{the flavor decomposition of the quark spin densities in the nucleon sea
      from measurements of $A_L$ in $W$ production,
      $\vec{p}p \rightarrow W + X \rightarrow e + X$,
}
\item{the quark transversity distributions from $A_{TT}$ 
      (i)   in di-jet production $\vec{p}\vec{p} \rightarrow \mathrm{di-jet} + X$,
            and
      (ii)  in $Z^0$ production
            $\vec{p}\vec{p} \rightarrow Z^0 \rightarrow e^+ + e^- +X$,
}
\end{enumerate}
in addition to higher twist effects in forward, inclusive $\pi^{0}$
production\cite{anselmino,sterman} and transfer of the beam polarization
into  the  final  state,  as  examined in reference\cite{vogelsang} for
$\Lambda$ production.

After  a short overview of the STAR detector systems crucial to the spin
measurements, the goal of this year's spin program is described in  some
detail,  followed  by a discussion  of future measurements of  the gluon
polarization and the flavor decomposition of the quark spin densities in
the nucleon sea.

\section{STAR detector}
The  STAR detector design facilitates the
identification   and   measurement  of  jets,  electrons,  photons,  and
neutral pions.

Its key component is a Time Projection Chamber (TPC)\cite{tpc} with full
azimuthal coverage, placed inside a uniform solenoidal 0.5\,T magnetic field.
The TPC thus provides tracking for charged particles with pseudorapidities
$|\eta|<1.5$, and is well suited for the reconstruction of hadronic jets.

Photons  and  electrons  are   identified using   a   lead-scintillator
Electromagnetic Calorimeter (EMC) with preshower layers, and measured to
an energy resolution $\Delta E/E \simeq 0.02 + 0.16/\sqrt{E}$.
The construction and integration of the EMC in STAR is proceeding in stages.
The modules installed to date (2001) cover
$0 < \eta <1$ and $\Delta \phi = 4\pi /5$.
Future additions will extend the coverage to $-1 < \eta < 2$ for the
full range in azimuth.
Each calorimeter  module  is  equipped  with a Shower Maximum
Detector (SMD) at a depth of about 5 radiation lengths, which provides  
the  high
spatial  resolution in the shower profile needed  to  resolve  
pairs of nearby photons characterizing $\pi^0$ decay.

The  last detector subsystem discussed here is the Forward Pion Detector
(FPD).
The  FPD consists of one prototype module of the EMC with its associated
SMD, and three similarly  shaped Pb-glass  detectors  positioned  around
the beampipe at a distance of 750\,cm from the interaction point.
Its primary purpose is the measurement of the transverse single spin
asymmetries in inclusive $\pi^0$ production.

Further  detail  on these and other STAR detector subsystems is provided
in reference\cite{tdr,fpd_proposal}.

\section{Single transverse spin asymmetries from 
$ \vec{p} + p \rightarrow \pi^0 + X$}

The  E704  collaboration  at FNAL has measured\cite{e704} 
the single transverse spin
asymmetries for the production of leading charged and neutral pions  up
to  large Feynman $x_F \simeq 2E_{\pi}/\sqrt{s}$, $0.0 < x_F < 0.8$, and
small  transverse  momenta  $0.5  <  p_T  <  2.0\,\mathrm{GeV}$  in $pp$
collisions at $\sqrt{s} = 20\,\mathrm{GeV}$.
The asymmetries are found to be large and are described by
models\cite{anselmino,sterman}  of  higher twist effects, which predict
that the asymmetries persist at RHIC energies
$\sqrt{s} = 200\,\mathrm{GeV}$.

\begin{figure}[ht] 
\begin{center}
\epsfig{file=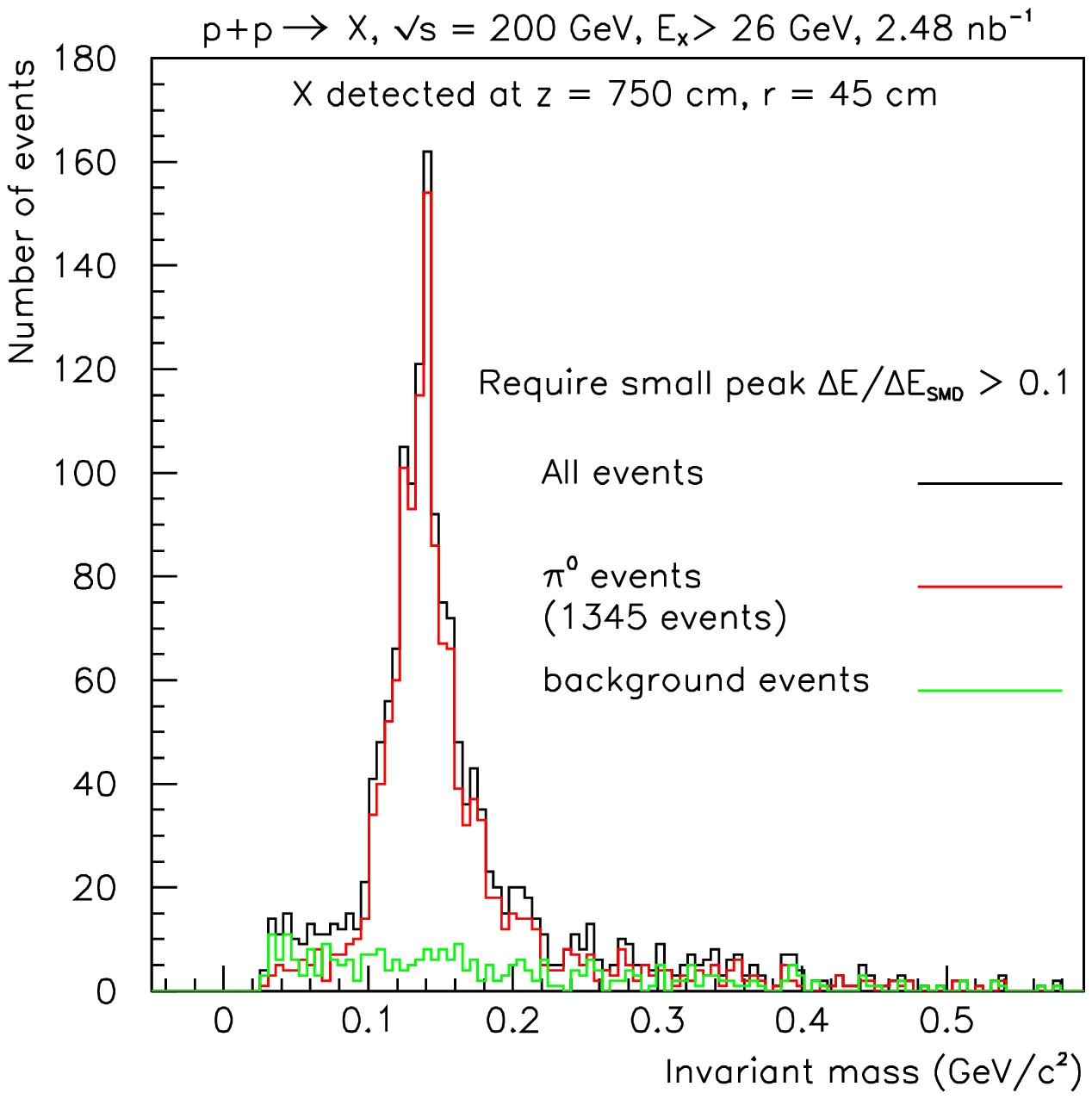,width=6.0cm}\hspace*{1.0cm}\epsfig{file=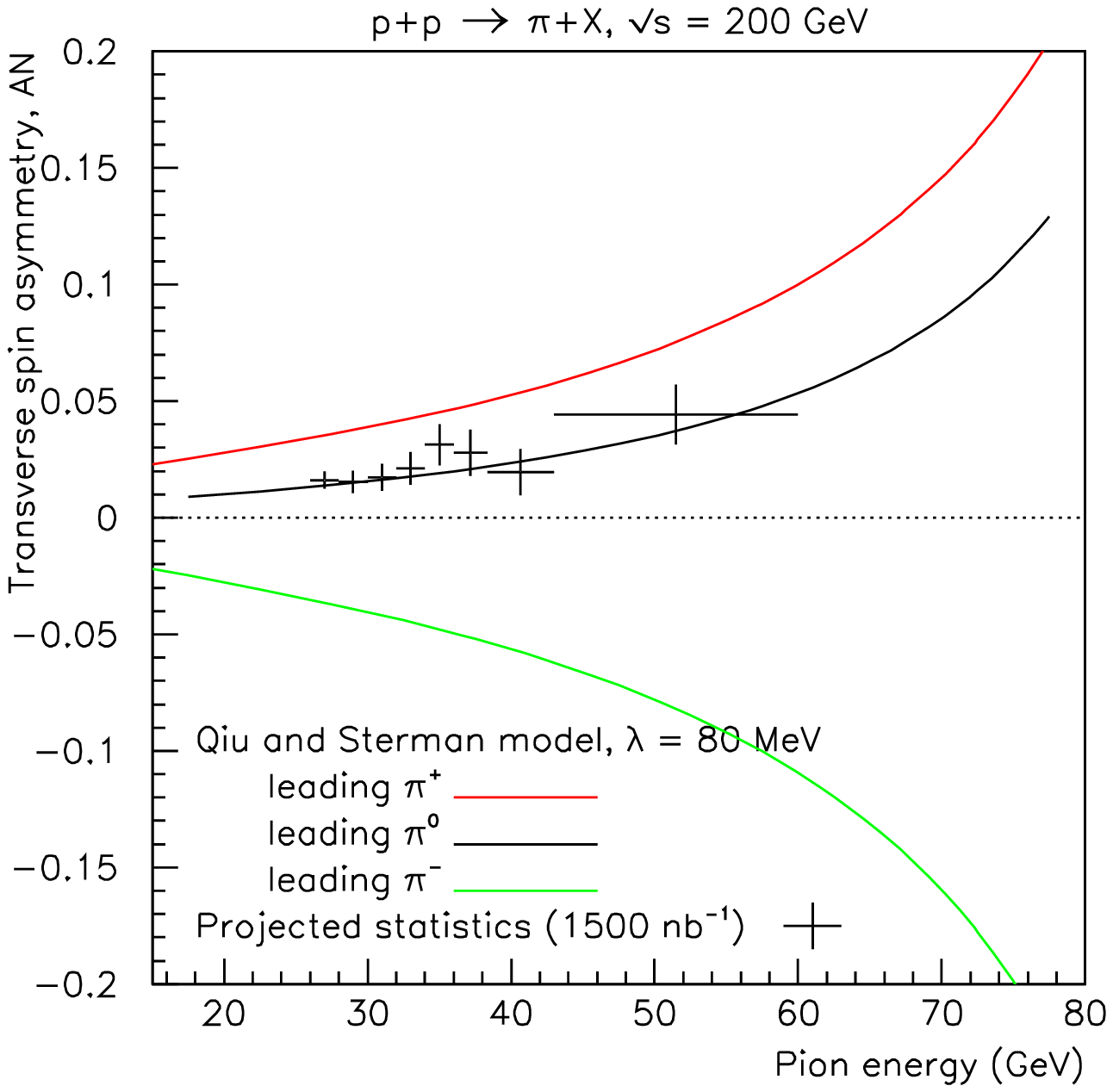,width=6.2cm}
\caption{
(Left) GEANT simulation of the di-photon invariant mass
distribution reconstructed with the STAR EMC in $pp$ collisions at
$\sqrt{s} = 200$ GeV simulated$^{13}$ with PYTHIA;
(Right) The single transverse asymmetry $A_N$ for   
$\vec{p} p \rightarrow \pi +X$ at $\sqrt{s}=200$ GeV as a function 
of the pion energy at a fixed transverse momentum of 
$1.5$ GeV. The lines are theoretical predictions
for neutral pion (continuous line) and charged pions (dashed lines).
The expected statistical precision of the 2001 STAR measurements
with the Forward neutral Pion Detector are indicated by the data points$^{13}$.
}
\end{center}
\label{inv}
\end{figure}

The  STAR  collaboration aims to measure the single transverse asymmetry
for neutral pions using the FPD, described above, for $0.2 < x_F <  0.6$
and $1 < p_T < 4\,\mathrm{GeV}$ during this year's spin 
running\cite{fpd_proposal}.
The  SMD  of  the  FPD  allows  the  reconstruction of the opening angle
$\phi_{\gamma_1 \gamma_2}$ between the decay   photons,  as  well  as  a
crude  partition  of  the  energy   of  the  photon pair measured in the
FPD prototype EMC module. The invariant mass
$M_{inv}^2=2E_{\gamma_1}E_{\gamma_2}(1-\cos{\phi_{\gamma_1 \gamma_2}})$
thus reconstructed is shown in Figure~\ref{inv} (left side).

The raw trigger rate from the FPD for the nominal luminosity of
$5 \times 10^{30}\,\mathrm{cm}^{-2}\mathrm{s}^{-1}$ is estimated from
PYTHIA\cite{pythia} simulations to be about 10\,Hz for an energy
threshold of 26\,GeV. Figure~\ref{inv} (right side) 
shows the projected data for 50\% beam polarization
from one week of running with regular data taking 
conditions.
The continuous (dashed) line shows the model\cite{sterman} asymmetry for
neutral (charged) pions.
Although the asymmetries for charged pions are expected to be even larger
than the asymmetry for neutral pions,  the observable  asymmetry with the
FPD is largest for $\pi^0$ since $\pi^+$ and $\pi^-$ cannot be
distinguished and $A(\pi^+) \simeq - A(\pi^{-})$.

The measurement of the single transverse spin asymmetry in $\pi^0$
production is the main goal of this year's STAR spin physics program.
The measurement will be the first at $\sqrt{s} = 200\,\mathrm{GeV}$,
and will prove STAR's ability to measure spin asymmetries in the RHIC
environment.
If the asymmetry is confirmed to be large, the
$\vec{p}p \rightarrow \pi^0 + X$ process may provide an experimental
method to \emph{in-situ} determine the beam polarization vector at
the STAR interaction region\cite{fpd_proposal}.
\section{Gluon polarization from $\vec{p} + \vec{p} \rightarrow \gamma +$ jet $+ X$} 
Prompt  photon  production  has been the classical tool to determine the
unpolarized gluon distribution for moderate and large values of $x$.
At leading order, the photon production in $pp$ collisions is dominated
by the gluon Compton process $q+g \rightarrow \gamma+q$.
PYTHIA-based simulations\cite{pythia,les} show that only about 10\% of
the production originates from other processes, predominantly from the
annihilation $q + \bar{q} \rightarrow \gamma + g$.
Restricting ourselves to  the  Compton  process, the double longitudinal spin
asymmetry $A_{LL}$ can be written at leading order as:
\begin{equation}
A_{LL} \simeq {\frac{\Delta G (x_g,Q^2)}{G(x_g,Q^2)}} \times A_1^p (x_q,Q^2) 
\times \hat{a}_{LL}^{\rm Compton},
\label{all_dg}
\end{equation}
where the proton asymmetry    $A_1^p$    is    known    from   inclusive
DIS\cite{emc,smc,slac,desy}     and     the      partonic      asymmetry
$\hat{a}_{LL}^{\rm Compton}$ can be calculated in perturbative QCD.
The QCD scale $Q^2$ is on the order of the $p_T^2$ for the prompt photon,
and the fraction $x_q (x_g)$ of the hadron momentum carried by the quark
(gluon) can be reconstructed on an event-by-event basis when the photon
and the jet are detected in coincidence,
\begin{equation}
  x_g^\mathrm{recon}=min(x_1,x_2),\hspace{0.2in}
  x_q^\mathrm{recon}=max(x_1,x_2),
\end{equation}
\begin{equation}
  x_{1(2)}= {\frac{p_{T,\gamma}}{\sqrt{s}}} 
  \left[ \exp{(\pm \eta_{\gamma})} + \exp{(\pm \eta_{\rm{jet}})} \right],
\end{equation}
where  $\eta_{\gamma}$  and  $\eta_\mathrm{jet}$  denote  the   observed
photon and jet pseudorapidities, $p_{T,\gamma}$ is the
observed transverse momentum of the photon, and
the jet energy needs not be determined.
This reconstruction works well for $x_q^\mathrm{recon} > 0.2$\cite{les}.
The measurement of $A_{LL}$ thus forms a direct determination of
$\Delta G(x)/G(x)$.
\begin{figure}[htbp] 
\begin{center}
\epsfig{file=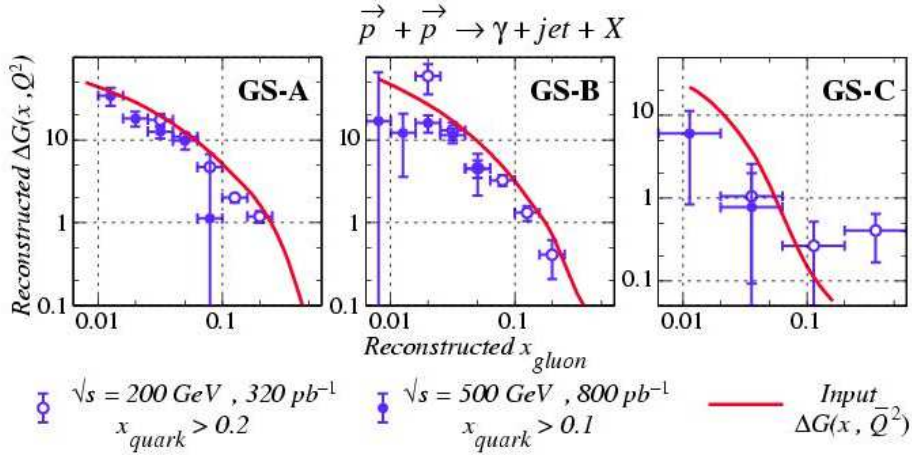,width=15.0cm}
  \caption{The estimated sensitivity of the future STAR measurements of
    $\Delta G(x)$ at an average $Q^2$ of 100\,GeV$^2$ from the
    $\vec{p}\vec{p} \rightarrow \gamma + \mathrm{jet} + X$ channel for
    three parametrizations of $\Delta G(x)$, as outlined in the text.
  }
  \label{prompt}
\end{center}
\end{figure}
To estimate the precision of the future STAR measurement,
three parametrizations from reference\cite{gs} (GS-A, GS-B, and GS-C)
of the largely unknown gluon distribution $\Delta G(x)$ were used.
At the initial scale $Q_0^2 = 4$ GeV$^2$, their integrals 
$\Delta G = \int_0^1 \Delta G(x) \mathrm{d}x$ evaluate to
$\Delta G (GS-A) = 1.7$, $\Delta G (GS-B) =  1.6$, and
$\Delta G (GS-C) = 1.0$.
The continuous lines in Figure~\ref{prompt} show the distributions
evolved to $Q^2=100\,\mathrm{GeV}^2$.

The data points show the values of $\Delta G$ reconstructed with
Eq.~(\ref{all_dg}) for integrated luminosities of \mbox{320\,pb$^{-1}$}
at \mbox{$\sqrt{s} = 200\,\mathrm{GeV}$} and
\mbox{800\,pb$^{-1}$} at \mbox{$\sqrt{s} = 500\,\mathrm{GeV}$}.
The measurement for two values of $\sqrt{s}$ is essential to cover
a relatively wide region in $x_g$, $0.01 < x_g < 0.3$.
The slight underestimation of the model values of $\Delta G$ results
mostly from
(i)  the neglect of the annihilation contribution to the observed
     asymmetry $A_{LL}$ in Eq.~(\ref{all_dg}), and
(ii) a small sample of analyzed events with incorrectly reconstructed
     kinematics
     ($x_q^\mathrm{recon} \neq x_q$ and $x_g^\mathrm{recon} \neq x_g$),
and this may be corrected for by simulation\cite{les}.
Backgrounds, predominantly high $p_T$ fragmentation photons and photons
from $\pi^0, \eta^0$ decays, potentially dilute $A_{LL}$.
These  backgrounds will be suppressed by requiring a large radius for
the isolation cone, and  by identifying  and rejecting pairs of decay 
photons with the SMD.

The integral $\Delta G = \int_{0}^{1} \Delta G(x)$ d$x$ from the STAR
measurements of $A_{LL}$ in
$\vec{p} \vec{p} \rightarrow \gamma + {\rm jet} + X$
is expected to be determined to a precision better than $\pm 0.5$.  
The knowledge of $\Delta G$ may be further improved by combining this
result with the results from  less favorable channels and  from other
experiments.

\section{Single spin asymmetries in $W$ production}

STAR  aims  to  decompose the quark spin densities in the nucleon sea by
measuring the parity violating single spin asymmetries for $W$
production in $\vec{p} + p \rightarrow W + X \rightarrow e + X$
collisions at $\sqrt{s} = 500\,\mathrm{GeV}$.
At these energies, the $W$  is produced predominantly through $q\bar{q}$
annihilation --- a valence-sea process  in  $pp$  collisions  ---  which
makes the $W$ an ideal probe.

The leading order $W$ production asymmetries read for
$Q^2 = M_W^2$\cite{alw}:
\begin{equation}
  A_L^{W^+}={{ \Delta u(x_1) \bar{d} (x_2) - \Delta \bar{d} (x_1) u(x_2) }
            \over{ u(x_1) \bar{d} (x_2) + \bar{d} (x_1) u(x_2)  }},
  \hspace{0.1in}
  A_L^{W^-}={{ \Delta d(x_1) \bar{u} (x_2) - \Delta \bar{u} (x_1) d(x_2)}
            \over{ d(x_1) \bar{u} (x_2) + \bar{u} (x_1) d(x_2) }},
  \label{alyw}
\end{equation}
where  the  transverse  momentum  of  the  $W$  is neglected and the $W$
production process is assumed to be dominated by   the  annihilation  of
light quarks\cite{mrst_lhc}.
The  fractions  $x_{1,2}$  of  the hadron momentum carried by the struck
quarks are related to the rapidity $y_W$ of the $W$ according
to~\cite{Peskin:1995ev}
\begin{equation}
  x_1 = {{M_W} \over{\sqrt{s}}} \exp{(y_W)} \hspace*{0.5cm} {\rm{and}}
  \hspace*{0.5cm}
  x_2 =  {{M_W} \over{ \sqrt{s}}}  \exp{(-y_W)},
\end{equation}
which thus needs to be reconstructed.
The rapidity $y_W$ can be determined from the observed rapidity
$y_e$ of the electron\cite{rhic_spin}.

The top part of Figure~3 shows evaluations of the asymmetry
based on the leading order polarized\cite{grvs2000} and unpolarized\cite{grv98}
parton distributions.
The  continuous  lines  correspond to the
'Standard Scenario'\cite{grvs2000} of
flavor symmetric light sea quark and antiquark
distributions at the initial QCD scale.

The dashed lines correspond to the 'Valence Scenario'\cite{grvs2000},
in which $SU(3)$ flavor symmetry is broken and the light sea densities
are different ($\Delta \bar{u} \ne \Delta \bar{d} \ne \Delta \bar{s}$).
The bottom part of Figure~3 compares the 'Standard Scenario'
asymmetries with their intuitive, approximate forms\cite{alw}:
\begin{equation}
  \begin{array}{lcr@{\hspace{0.2in}}lcr@{\hspace{0.3in}}lr}
    A_L^{W^+} & \simeq & - {{\Delta \bar{d} } \over { \bar{d} }}, &
    A_L^{W^-} & \simeq & - {{\Delta \bar{u} } \over { \bar{u} }}, &
    \mbox{for $y_W \leq$} & -1,  \\
    A_L^{W^+} & \simeq & {{\Delta u } \over { u }}, &
    A_L^{W^-} & \simeq & {{\Delta d } \over { d }}, &
    \mbox{for $y_W \geq$} & 1.
  \end{array}
\end{equation}

\begin{figure}[htbp]
\vspace*{-0.8cm} 
\begin{center}
%\label{wfig}
\epsfig{file=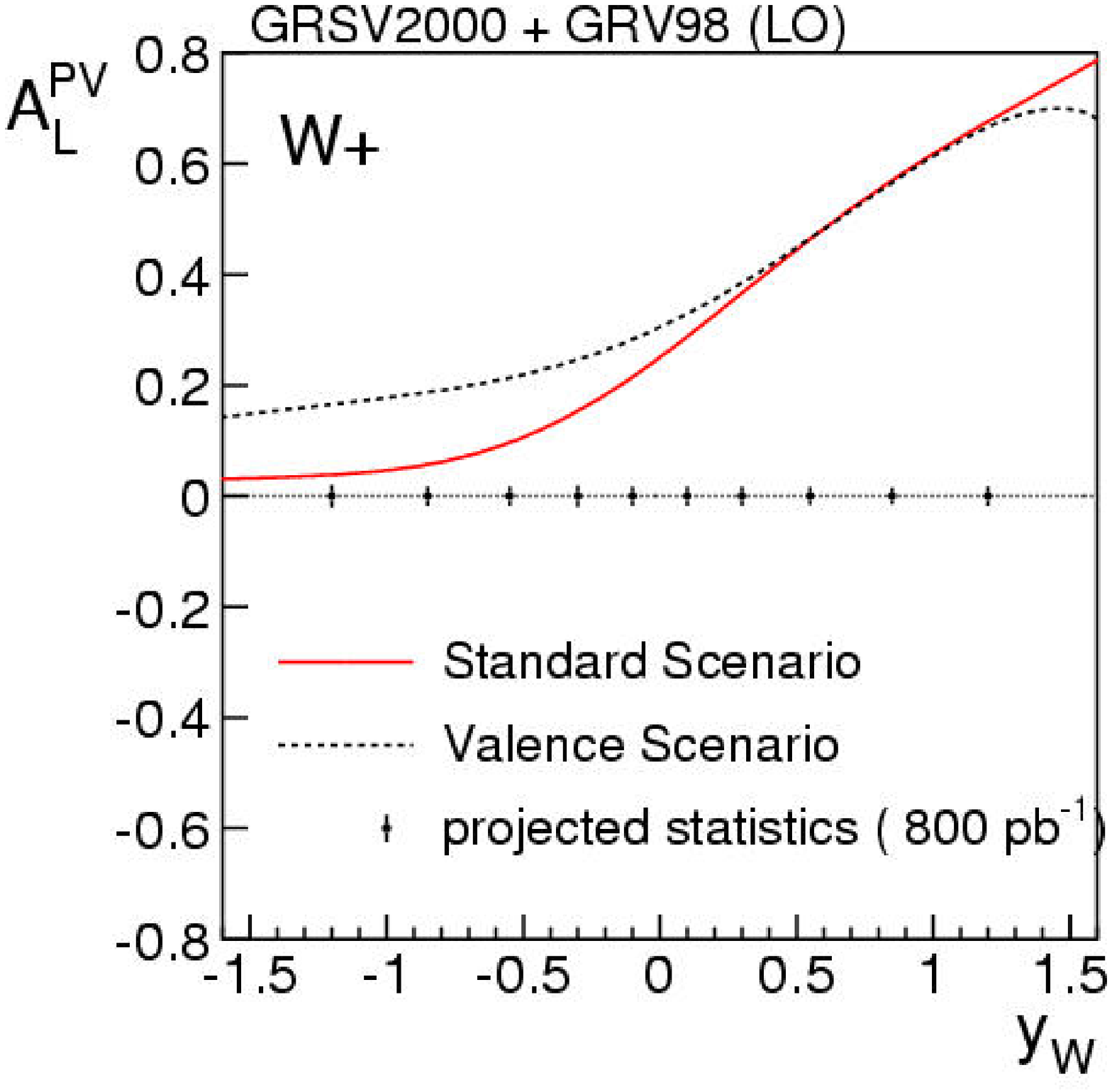,width=7.5cm}\epsfig{file=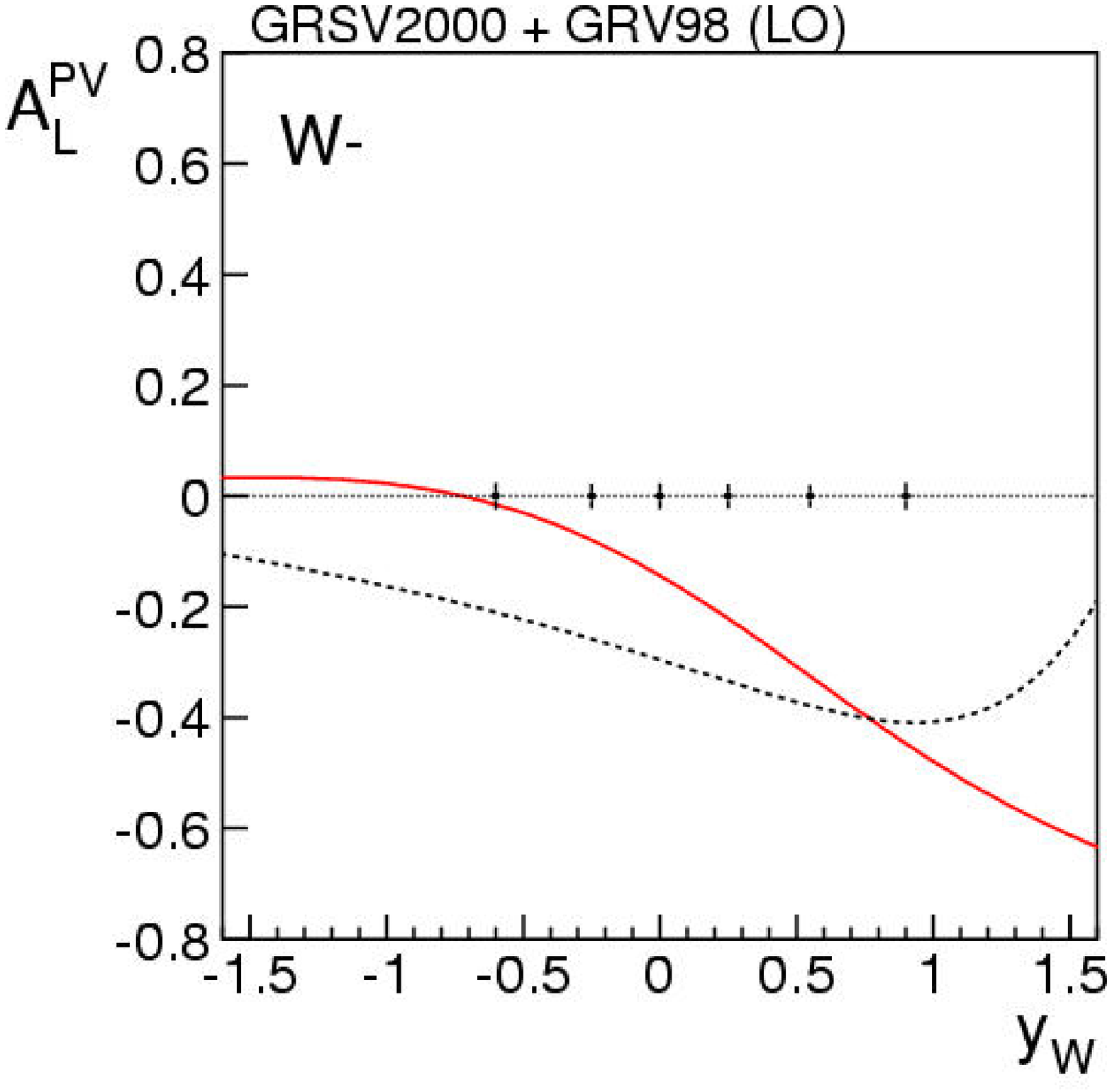,width=7.5cm} \\
\vspace*{-2.5cm}
\epsfig{file=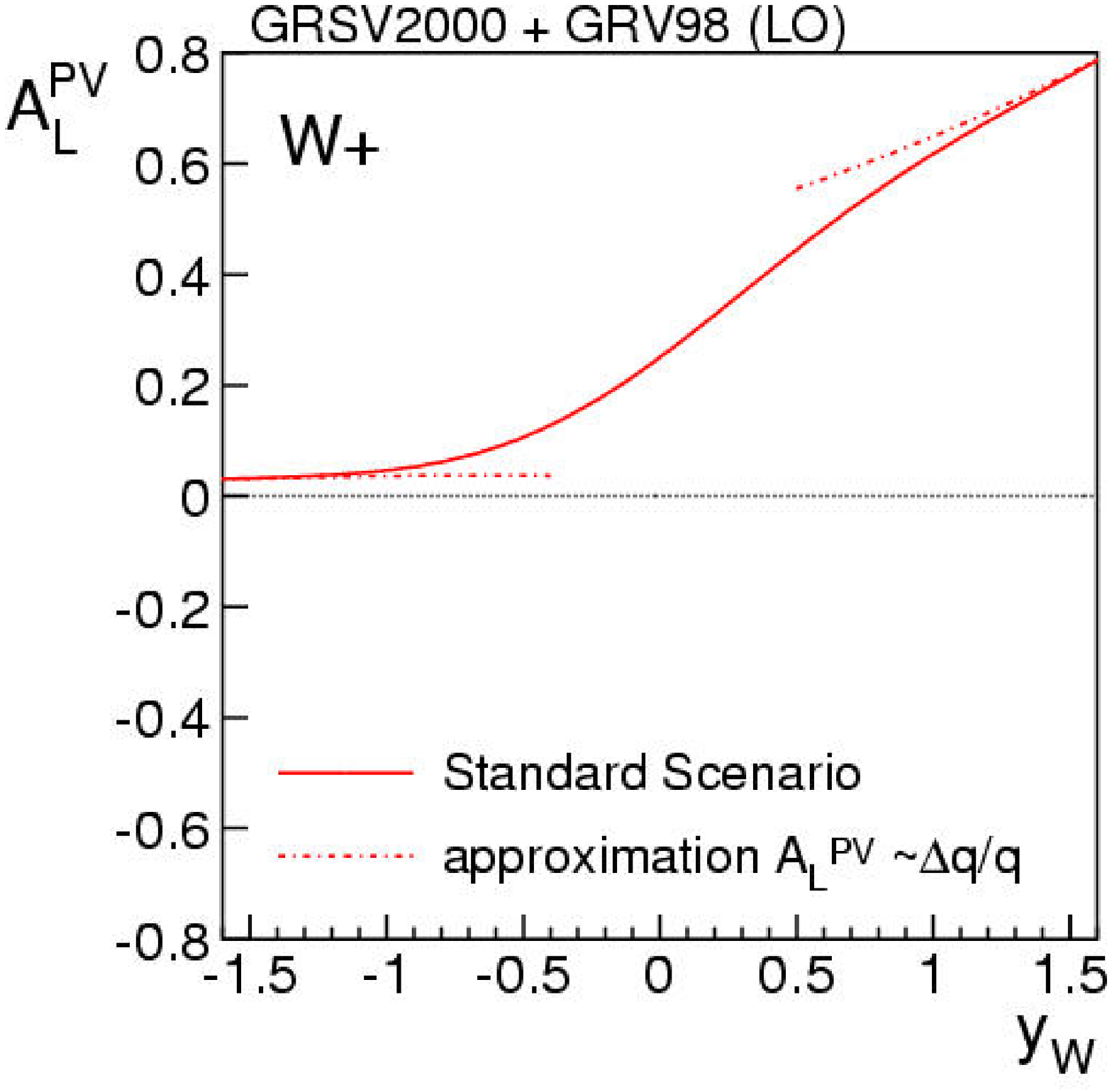,width=7.5cm}\epsfig{file=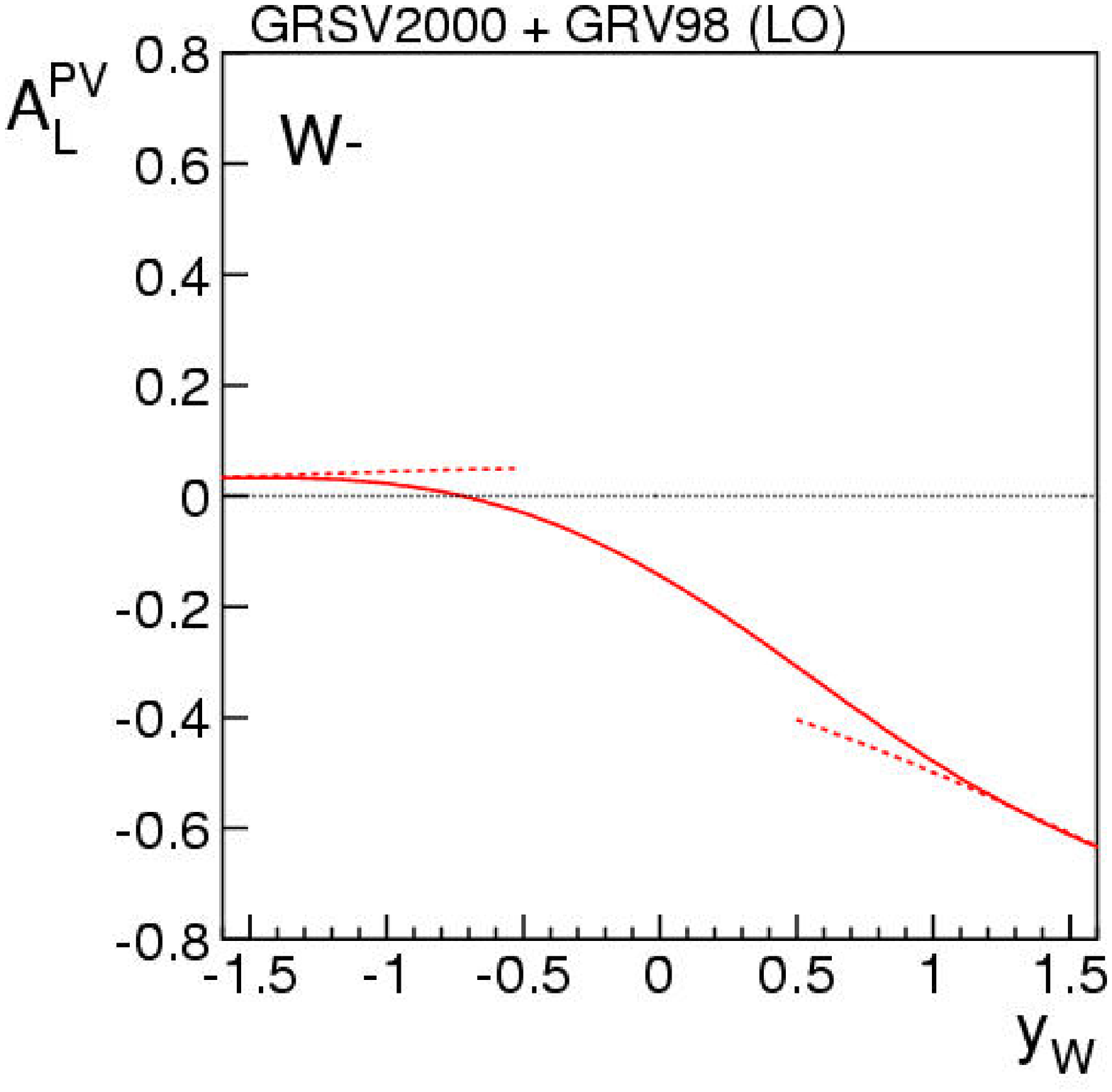,width=7.5cm} 
\vspace*{-1.5cm}
\caption{Top half: $A_L^W (y_W)$ for $W^+$ (left) and for $W^-$ (right) for
         a flavor symmetric distribution of the light sea quark and antiquark
         polarizations (continuous line) and for a flavor asymmetric
         distribution (dashed line).
         The points indicate the projected precision of the STAR measurements
         for 800\,pb$^{-1}$ at $\sqrt{s} = 500\,\mathrm{GeV}$.
         Bottom half: the asymmetries $A_L^W (y_W)$ for a flavor symmetric
         distribution of the light sea quark and antiquark polarizations
         (continuous lines) together with the intuitive approximations
         $A_L^W \simeq \Delta q / q$ (dashed lines), cf. Eq.(7).
}
\end{center}
\end{figure}

The points in Figure~3 show the projected statistical precision
of future STAR measurements at $\sqrt{s} = 500\,\mathrm{GeV}$.
These projections are based on an integrated luminosity of
\mbox{800\,pb$^{-1}$} and account for preliminary data selections
used to reliably relate the rapidities $y_e$ and $y_W$.
Potentially large hadronic backgrounds will be rejected with the
Preshower and SMD subsystems of the EMC, whereas backgrounds from
heavy quark and $Z^0$ leptonic decays are estimated to be manageably
small. The measurements -- while demanding -- are thus expected to verify
various symmetry scenarios and to aid the flavor decomposition of quark
spin densities in the nucleon sea.

\section{Summary}

The STAR  collaboration is about to start its spin physics measurements,
and aims in its initial running period  of  about  five  weeks  for  the
measurement  of  the single transverse spin asymmetry  in the production
of  neutral  pions,  using  a   recently   installed   and  successfully
commissioned Forward neutral Pion Detector.

During future running periods the aim of the spin measurements is to determine
the polarization
$\Delta G$ of gluons in the  nucleon,  the  flavor  composition  of  the
nucleon sea, and the yet unexplored quark transversity distributions.
STAR's unique ability to measure $\Delta G$ through coincident detection
of direct photons and jets from $\vec{p}\vec{p}$ collisions is  expected
to yield  a  particularly competitive  measurement  of $\Delta G$ over a
wide kinematic range, $0.01 < x_g < 0.3$.
Furthermore, the high beam energies at RHIC open a window  to study the
flavor composition of quarks in the nucleon through weak-interactions.

RHIC starts colliding polarized protons in December 2001, thereby opening
a new era in high energy spin physics --- stay tuned!

\end{document}